\documentclass{mn2e}
\usepackage{graphicx}
\usepackage{txfonts}

\newcommand{\degree}{\ensuremath{\mathrm{^\circ}}}
\newcommand{\arcm}{\ensuremath{\mathrm{^\prime}\;}}
\newcommand{\arcs}{\ensuremath{\arcmm\hskip -0.1em\arcmm \;}}
\newcommand{\arcmm}{\ensuremath{\mathrm{^\prime}}}

   \title[A giant diffuse non-thermal source in A3411 and A3412]
          {The nature of the giant diffuse non-thermal source in the 
           A3411--A3412 complex}

   \author[G. Giovannini et al.]{G. Giovannini$^{1,2}$\thanks{E-mail:ggiovann@ira.inaf.it},
V. Vacca$^{1,2}$, M. Girardi$^{3}$, L. Feretti$^{2}$, 
           F. Govoni$^{4}$, M. Murgia$^{4}$
\\
$^{1}$Dipartimento di Fisica e Astronomia, via Ranzani 1, 40127 Bologna, I \\ 
$^{2}$Istituto di Radioastronomia-INAF, via P.Gobetti 101, 40129 Bologna, Italy \\
$^{3}$Dipartimento di Fisica-Sezione di Astronomia, via Tiepolo 11, 34143 Trieste, Italy \\
$^{4}$Osservatorio Astronomico di Cagliari-INAF, Strada 54, Loc. Poggio dei Pini, 09012 Capoterra (Ca), Italy}

\begin{document}

\date{ }

\pagerange{\pageref{firstpage}--\pageref{lastpage}} \pubyear{2013}

\maketitle

\label{firstpage}

\begin{abstract}
VLA deep radio images at 1.4 GHz in total intensity 
and polarization reveal a diffuse non-thermal source in the interacting
clusters A3411 -- A3412. 
Moreover
a small-size low power radio halo at the center of the merging cluster 
A3411 is found.
We present here new optical and X-ray data 
and discuss the nature and properties of the diffuse non-thermal source.
We suggest that the giant diffuse radio source is 
related  to the presence of a large scale filamentary structure and to 
multiple mergers in the A3411--A3412 complex.

\end{abstract}

\begin{keywords}
Galaxies:cluster:non-thermal -- Clusters: individual: 
Abell 3411, Abell 3412 -- Cosmology: large-scale structure of the Universe
\end{keywords}

%

\section{Introduction}

Clusters of galaxies are characterized by X-ray emission from a hot 
intra-cluster medium (ICM, $T \sim 2{-}10$ keV). Thermal emission is a common 
property of all clusters of galaxies and has been detected even in poor 
groups as well as in optical filaments connecting rich clusters.

In some clusters the ICM is also characterized by non-thermal diffuse 
emission: giant radio sources with a spatial extent similar to that of the hot 
ICM, which are called radio halos or relics, according to their morphology
and location in the cluster (see \cite{fer12}, 
for a recent review). 
These sources are not directly associated
with the activity of individual galaxies and are related to physical 
properties of the whole cluster. 
In a few cases a diffuse non thermal emission is found also in 
filamentary structures (see e.g., \cite{gio10}), and 
in A399 \& A401 we found two radio halos in two interacting clusters 
(\cite{mur10}).

In the last years, our knowledge of the non-thermal emission in galaxy clusters
and the properties of large scale magnetic fields largely increased thanks
to the discoveries of new halo and relic sources 
and the large development in simulations and theoretical models (see e.g.,
\cite{fer12}, \cite{gov13}, \cite{ven13}, \cite{don13}, 
\cite{bon12}, \cite{vwe12}). 

There is a substantial evidence that radio halos are found in 
clusters showing significant substructures in X-ray images and
complex gas temperature distributions, both signatures of cluster mergers 
(e.g. \cite{fer99}, \cite{gov04}, \cite{gio09}, \cite{cas10}, 
\cite{fer12}). 
The monochromatic radio power of a halo at 1.4 GHz correlates with the cluster 
X-ray luminosity, mass and temperature (e.g. \cite{fer00}, \cite{bru07}, 
\cite{gio09}), implying a direct 
connection between
the radio and X-ray plasmas (\cite{fer00}; \cite{gov01}).

In addition, diffuse radio sources classified as radio relics are
detected in cluster peripheral regions (see e.g. \cite{gio04},
\cite{vwe11}, \cite{fer12}).
Observations of relics provide the best indication for the presence of 
magnetic fields and relativistic particles in cluster outskirts. 
Relic sources are strongly polarized ($\sim$ 20--30 \%), 
and typically show an elongated radio structure, with the major axis 
roughly perpendicular to the direction of the cluster center. 
When observed with high angular resolution, they generally show an asymmetric 
transverse profile, with a sharp edge usually on the side toward the cluster 
outer edge. These morphologies are in very good agreement with models 
predicting that these sources are related to large-scale shocks generated 
during cluster merger events (see e.g. \cite{bru11}).
However, \cite{ogre13}, discussing the 
{\it Toothbrush} cluster, found several difficulties in
this interpretation.
Moreover we note the existence of
relic sources with roundish structures, 
classified in  \cite{fer12}, whose morhology is quite different
from that expected if they were originated by shock waves.

In a few cases, non-thermal emissions has been detected in structures
connecting merging clusters (filaments), see e.g., 
the filament surrounding
the cluster ZW2341.1+0000 (\cite{bag02} and \cite{gio10}) 
and the bridge of radio
emission extending about 1 Mpc in size, connecting
the central halo Coma C and the peripheral relic source 1253+275 in the Coma 
Cluster (\cite{kim89}). The orientation of this feature is, suggestively, 
aligned with the 
direction to A1367 (see also \cite{bro11}).

Recently \cite{vwe13} presented the detection of a new halo source 
located at the center of the cluster A3411, and a complex region
of diffuse, polarized emission found in the southeastern outskirts of A3411, 
along the projected
merger axis of the system. They classify this region as a peculiar 
1.9 Mpc scale radio relic. 

Here we present new optical results on this complex region, and
the large scale X-ray image from archive ROSAT data, both showing 
that A3411 and A3412
are two interacting clusters in a large scale filament.
These data compared with
high and low resolution polarized VLA images, show 
that the giant diffuse radio source is likely related  to the filamentary
structure and to the presence of multiple mergers in the A3411--A3412 complex.

The intrinsic parameters quoted in this paper are computed for
a $\Lambda$CDM cosmology with $H_0$ = 71 km s$^{-1}$Mpc$^{-1}$,
$\Omega_m$ = 0.27, and $\Omega_{\Lambda}$ = 0.73.
At z =0.1687, the luminosity distance is 803 Mpc, and the angular conversion 
factor is 2.85 kpc/arcsec.

\section{Optical data}

A3411 and A3412 are on the galactic plane therefore available X-ray and optical
information is poor. The cluster centers as reported in the 
Nasa/IPAC Extragalactic Database (NED) are: 
RA $08h:41m:47.7s$ and DEC
$-17^{\circ}:28^{\prime}:46^{\prime\prime}$ for A3411,
and RA $08h:42m:05.6s$ and DEC
$-17^{\circ}:35^{\prime}:47^{\prime\prime}$ for A3412.

\begin{figure} 
\includegraphics[width=80mm]{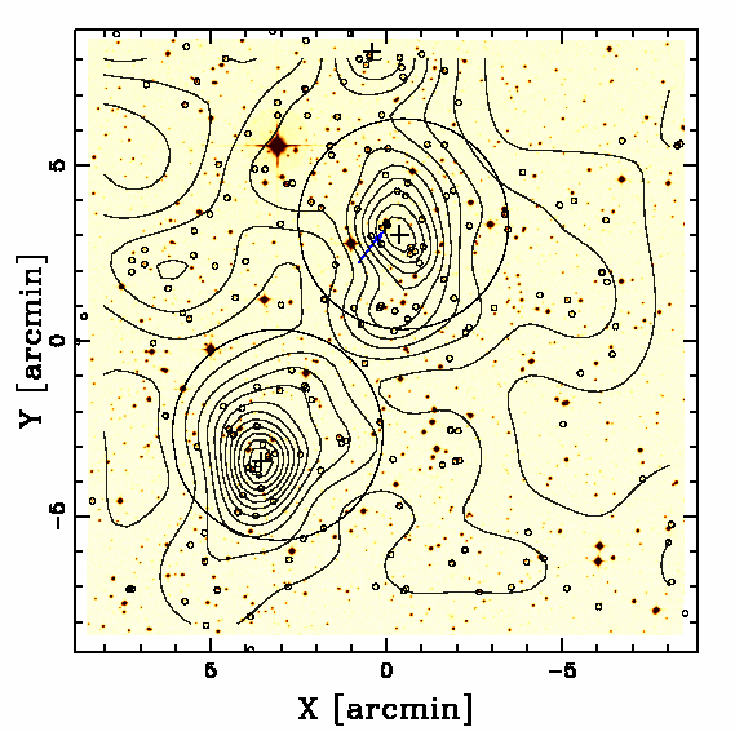} 
\caption {DSS
Spatial distribution on the sky of likely cluster members -- very
small circles -- and relative isodensity contour map are superimposed
on the DSS2-red image.  
Levels are: 0.25 0.50 0.75 ... 3.00 gals/$arcmin^2$.
The center and the size ($\sim 3$ Mpc)
correspond to those of Fig. \ref{fig:2}  for an easy comparison. Large circles
indicate the 3\arcmm-radius regions around A3411 (Northern structure) and 
A3412 (Southern structure) centered
on the X-ray peaks as listed in the text. Crosses highlight the positions
of the peaks in the galaxy distribution as detected in our 2D-DEDICA
analysis. The arrow near to the A3411 center indicates the galaxy with known 
redshift
listed in the 6dF Galaxy Survey, very close to a star.}
\label{fig:1}
\end{figure}

\begin{figure}
\includegraphics[width=8.5cm]{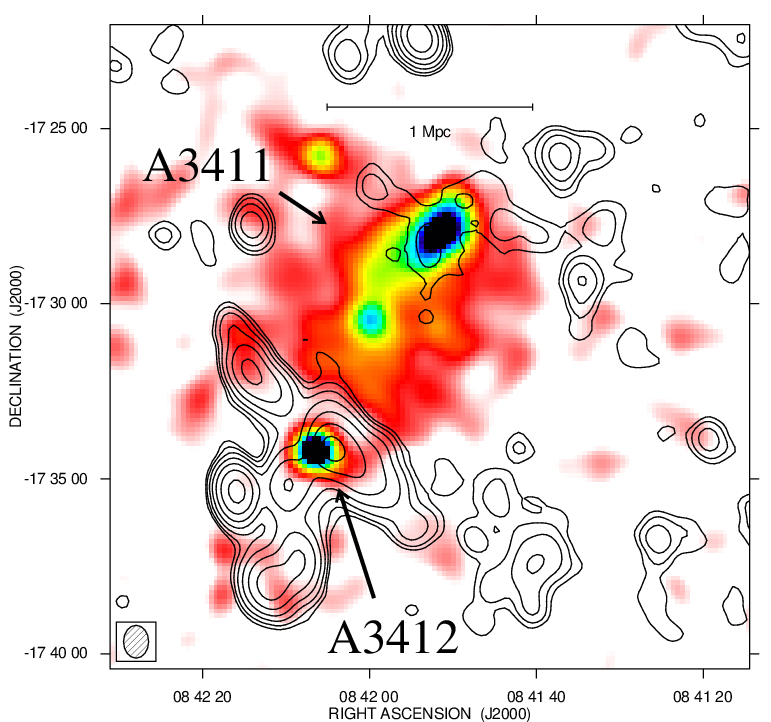}
\caption{Radio X-ray overlay of the A3411 -- A3412 region.
Contours show the radio source at 1.4 
GHz. The HPBW is 56.3'' $\times$ 43.1'' (PA -17$^\circ$).
Levels are: 0.12 0.24 0.48... mJy/beam. The noise level is 0.04 mJy/beam. 
In colour the X-ray image from archive ROSAT data is shown.}
\label{fig:2}
\end{figure}

For A3411 (see also \cite{vwe13}) it is reported a redshift = 0.1687 
(\cite{ebe02}), and the 6dF Galaxy Survey (\cite{jon09}) lists a 2MASS
galaxy at $z=0.162569$, very close to the A3411 center 
(see Fig.~\ref{fig:1}).  No redshift is reported for A3412.

The distance between the two cluster centers is $\sim$ 8.2\arcm 
corresponding to $\sim$ 1.4 Mpc assuming
the redshift of A3411.

To analyze the structure of the A3411--A3412 complex we constructed a
photometric catalog in a large region of 30\arcmm-radius around the
A3411 center extracting the objects classified as ``galaxies'' and
having both $B_j$ and $R$ magnitudes available and having $R<20$ mag
from the SSA - SuperCOSMOS Science 
Archive\footnote{http://surveys.roe.ac.uk/ssa/}.  In this
photometric catalog, we selected ``likely cluster members'' on the basis
of the color--magnitude relation (hereafter CMR), which indicates the
early-type galaxy locus.  We considered galaxies having observed
colors within 0.4 mag from $B_j-R=1.2$ where 1.2 is roughly the
expected color for CMR galaxies (considering typical $B-R$ values at
$z=0.17$ in \cite{lop04}, opportune magnitude conversions
and Galactic absorption). Note that we prefer to rely on the
expected CMR relation and a quite large color range since galaxy
colors (and star/galaxy separation too) are expected to have large
uncertainties in the A3411 region.

Figure~\ref{fig:1} 
shows a zoomed region of the contour map for the
likely members (860 galaxies within the whole $\sim$ 30\arcm region) as
obtained through to the 2D adaptive--kernel method (2D DEDICA,
\cite{pis96}). The two densest peaks in the galaxy distribution
correspond to A3411 and A3412, nominally being A3412 even denser than
A3411. A third significant, but much less dense, peak lies at north at
the limit of the plot we show.  
The three density peaks are significant at the $>99.99\%$ c.l. and
Table~\ref{tabdedica2d} lists relevant information: the number of
assigned members, $N_{\rm S}$ (Col.~2); the peak position (Col.~3);
the density (relative to the densest peak), $\rho_{\rm S}$ (Col.~4);
the value of $\chi^2$ for each peak, $\chi ^2_{\rm S}$ (Col.~5).

\begin{table}
        \caption[]{2D DEDICA optical structure}
         \label{tabdedica2d}
            $$
         \begin{array}{l r c c c }
            \hline
            \noalign{\smallskip}
            \hline
            \noalign{\smallskip}
\mathrm{Subclump} & N_{\rm S} & \alpha({\rm J}2000),\,\delta({\rm J}2000)&\rho_{
\rm S}&\chi^2_{\rm S}\\
& & \mathrm{h:m:s,\degree:\arcmm:\arcs}&&\\
         \hline
         \noalign{\smallskip}
\mathrm{A3412}          & 49&08\ 42\ 08.0-17\ 34\ 45&1.00&21\\
\mathrm{A3411}          & 34&08\ 41\ 51.4-17\ 28\ 18&0.83&16\\
\mathrm{Northern peak}  & 50&08\ 41\ 54.7-17\ 23\ 05&0.55&12\\
              \noalign{\smallskip}
              \noalign{\smallskip}
            \hline
            \noalign{\smallskip}
            \hline
         \end{array}
$$
         \end{table}

To obtain more reliable positions and relative densities, we stress the
need of better quality images covering the whole cluster complex in
two magnitude bands in a homogeneus way, while to-date available, more
recent images do not have both the advantages.

The relatively small distance between A3411 and A3412 and the presence of 
galaxies in the connecting 
region suggests that they are two interacting clusters.
\cite{vwe13} classify the central galaxy in A3411 as a cD galaxy
from a VLT FORS1 image of the central part of the cluster.

\section{X-ray data}

In Fig. \ref{fig:2} the ROSAT image obtained from archive data (observation 
request 801009)
of this region is shown in color. 
The image shows an extended complex emission with 
three aligned major clumps.

We estimated the X-ray luminosity of the whole region, from ROSAT All Sky
Survey data,
by considering a weighted averaged total Galactic HI column density
4.67 $\times$ 10$^{20}$ cm$^{-2}$ from the Leiden/Argentine/Bonn (LAB) Survey 
(\cite{kal05}), an APEC model with kT=5 keV and metallicity 
0.4$Z_{\odot}$.
We measured in the $0.1-2.4$\,keV band a total X-ray
luminosity of $\sim5.3\times10^{44}$\,erg/s, over a region of
$\sim13^{\prime}$, centered at the location of the less bright X-ray peak
and including all the three clumps.
This value is slightly higher, but in agreement within the errors
with the value published in \cite{ebe02}.

In more detail we note that the X-ray image shows an extended emission
with a peak near to the A3411 optical position: 
RA $08h:41m:51s$ and DEC
$-17^{\circ}:28^{\prime}:00^{\prime\prime}$.
We assume this as the A3411 cluster center. The X-ray
emission is extended in S direction and a secondary peak is
present at RA $08h:42m:00s$ DEC $-17^{\circ}:30^{\prime}:30^{\prime\prime}$.
For the A3411 structure, a total X-ray luminosity of
$\sim5.0\times10^{44}$\,erg/s is measured over a region of
$\sim7.5^{\prime}$.
From X-ray data we conclude that A3411 is a bright
X-ray cluster in a merging phase.

At RA $08h:42m:06s$ DEC $-17^{\circ}:34^{\prime}:00^{\prime\prime}$ one
more bright X-ray emission is present, slightly extended to SW.
This compact structure has
a X-ray luminosity of $\sim 3.2\times 10^{43}$\,erg/s over a region of
$\sim 4^{\prime}$ in size.
Because of its position we identify this X-ray emission with A3412,
even if its compactness could suggest the identification of
the X-ray structure with a discrete source (e.g., a radio galaxy; 
see Sect. 4.1).
We note that A3412 is
aligned with the very elongated X-ray structure of A3411 and that the X-ray 
emission is in good agreement with the optical galaxy distribution. 

We compared ROSAT X-ray data with the Chandra X-ray image of A3411
published by \cite{vwe13} (their Fig. 1). Both images confirm the evidence 
that A3411
is a cluster in a merging phase. The high resolution Chandra
image by \cite{vwe13} (their Fig. 1-right) shows the inner disturbed region
of A3411.
The secondary peak in between A3411 and A3412, present at $\sim$ 6.5 
$\sigma$ level in the ROSAT All Sky Survey data,
is marginally visible being at the bottom left corner.

In the ROSAT image one more peak of emission is
present at the NE of A3411 at $\sim$ 4.7 $\sigma$ level. 
No radio emission is 
present from this
region and no optical condensation was found. With present data we cannot
confirm if it is one more substructure related to A3411 or if it is an 
unrelated discrete source. This structure is outside the Fig. 1-right
in \cite{vwe13}.
We note that the whole X-ray structure (including secondary peaks)
is present both in the X-ray image derived from the ROSAT All Sky Survey data 
and in the image from the ROSAT archive pointed observation: the two
independent images are in very good agreement.

\section{Radio data}

\subsection{Full Intensity  Images}

A field centered on A3411 was observed at 1.4 GHz in full polarization mode
in October 2003 in A/B configuration and in June 2004 in C/D configuration
(project AC696). Because of the small angular distance between A3411 and
A3412 ($\sim$ 7'), both clusters are well imaged with the present data.
 
Calibration and imaging were performed with the NRAO Astronomical
Image Processing System (AIPS). 
After the editing of bad points, self-calibration 
was applied to both datasets,
to remove residual phase and gain variations. The two datasets were
combined together to produce final images. The final good uv-coverage
of the combined dataset,
allowed us to derive images with a low noise level.
We obtained images at low angular resolution using the factor 
ROBUST = 5 in the IMAGR task, at intermediate resolution (15'', ROBUST = 1), 
and at high angular resolution (5'', ROBUST = -5). Final images
were corrected for the primary beam attenuation, to properly measure the
flux density.
Polarized images were obtained combining Q and U images and correcting for the
positive bias, using standard AIPS tasks.

\begin{figure} 
\includegraphics[width=80mm, angle = -90]{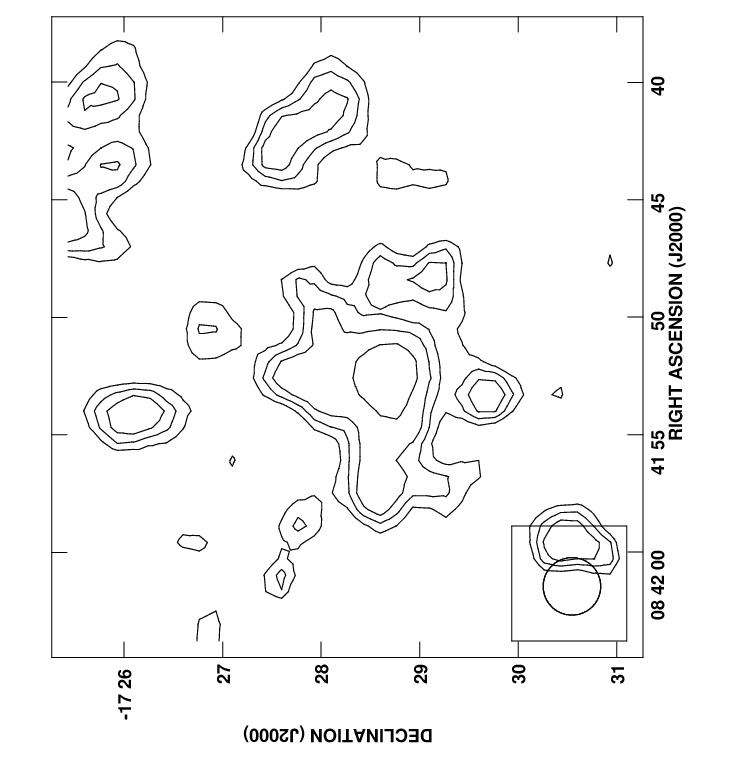} 
\caption {Contour image of the central diffuse radio source identified
as a radio halo at the center of A3411. Discrete sources in the halo
region have been subtracted.
The HPBW is 35''. Contours are 0.06 0.08 0.10 0.15 0.30 0.50 mJy/beam.
The noise level is 0.03 mJy/beam.}
\label{fig:fighalo}
\end{figure}
 \begin{figure}

 \centering
 \includegraphics[width=8.0cm, angle = -90]{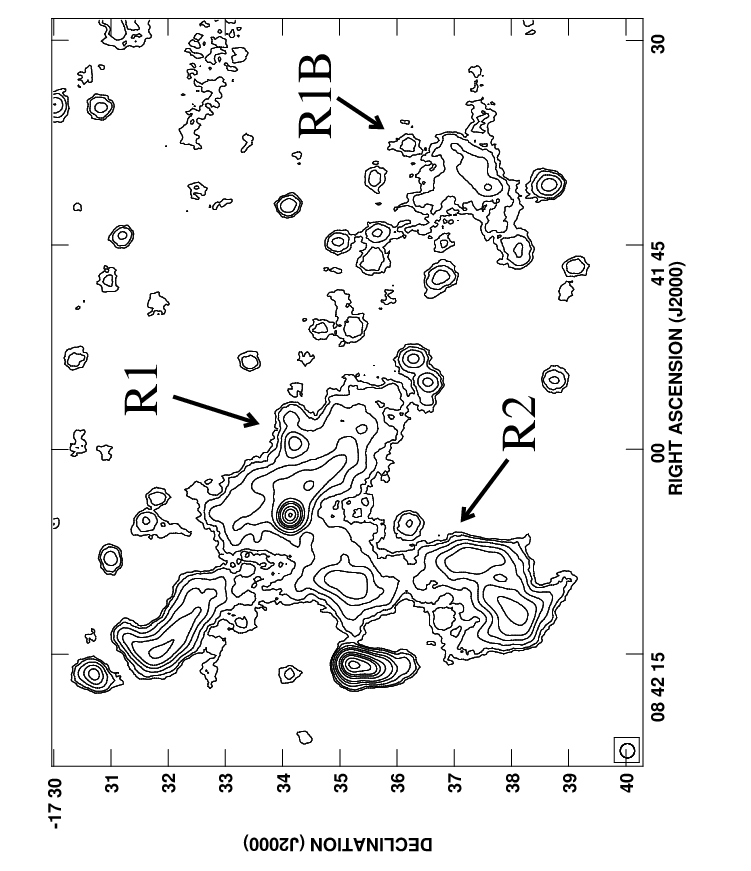}
    \caption{Contour image of the A3411--A3412 region. The HPBW is 
15 arcsec. Contours are: 0.04 0.07 0.15 0.3 0.5 1 1.5 2 3 4 5 mJy/beam;
the noise level is 0.02 mJy/beam. 
R1, R1B, and R2 refer to radio structures discussed in the text.}
       \label{fig:3}
 \end{figure}

 \begin{figure}
 \centering
 \includegraphics[width=8.0cm, angle = 0]{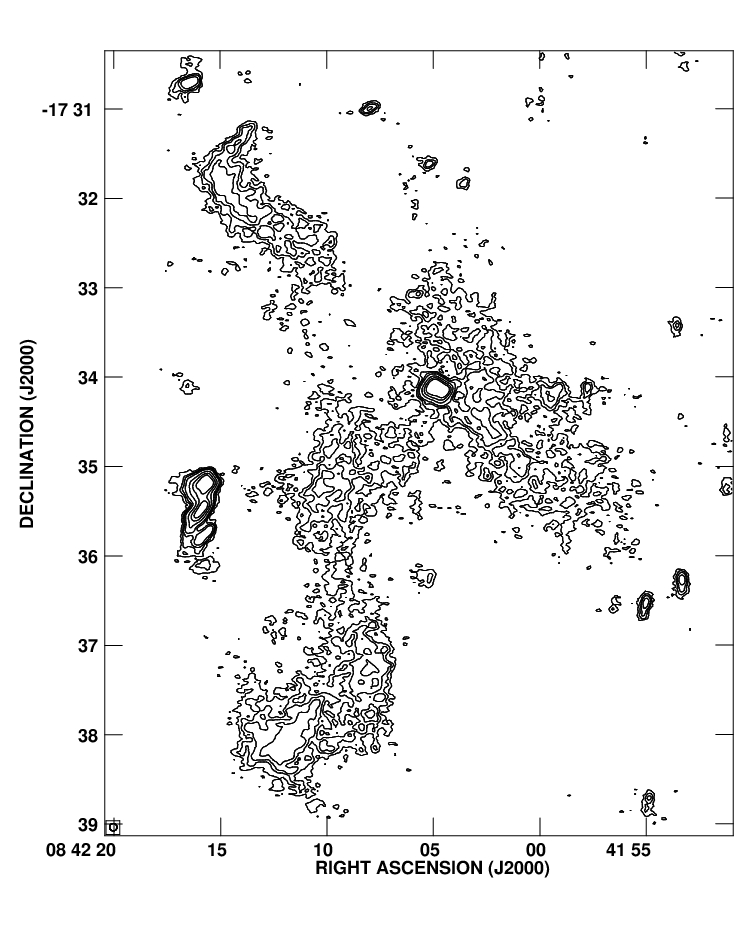}
    \caption{Contour map of the diffuse emission regions R1 and R2 (see text).
The HPBW is 5''. Levels are: 0.04 0.07 0.1 0.15 0.3 0.5 0.7 mJy/beam; the 
noise level is 0.016 mJy/beam.}
       \label{fig:4}
 \end{figure}

In Fig. \ref{fig:2}, we show the radio image at the angular resolution of 
56.3'' $\times$ 43.1'' in PA 17$^\circ$, overimposed onto a ROSAT X-ray image 
(colour)
obtained from archive data (Observing request number 801009).

At the position of A3411, coincident with the main peak of the X-ray emission
we confirm (\cite{vwe13}) the presence of a diffuse faint radio source 
that we interpret as a possible small size central halo (see 
Fig. \ref{fig:2}, and Fig. \ref{fig:fighalo}). 
The morphology
of this extended source is rather irregular in agreement with the 
recent finding of irregular low power radio halos,
discussed in \cite{gio09}. At high resolution 
a few contaminating discrete sources are present, subtracted in 
Fig. \ref{fig:fighalo} (see Sect. 5.2). 

Moreover as discussed by \cite{vwe13} a remarkable extended diffuse 
radio 
source is detected at the peripheral region of A3411. We confirm
the presence of this structure crossing the A3412 cluster 
center. For a more clear discussion, we separate
this extended diffuse, non-thermal source in two sub-structures: R1 
elongated perpendicularly to the 
A3411 and A3412 filament, and R2 a N-S more elongated structure 
aligned with the A3411 and A3412 filament, southern to A3412 (see 
Fig. \ref{fig:3}). 
To the West side of R1 another diffuse, non-thermal source is present with a 
lower surface brightness (named R1B in Fig. \ref{fig:3}), 
aligned with the brighter structure. Low resolution 
images show that these structures are connected and that the diffuse,
non-thermal source R1 extends up to $\sim$
RA $08h:41m:40s$ DEC $-17^{\circ}:37^{\prime}:30^{\prime\prime}$ 
with a total size $\sim$ 1.9 Mpc, in agreement with \cite{vwe13}.

The R2 diffuse structure starts from the A3412 region, 
and extends outwards, in the direction of the A3411-A3412 connecting line 
(Fig. \ref{fig:3}).
Its size is $\sim$ 860 kpc S to A3412.

Both radio structures if observed at high resolution (Fig. \ref{fig:4}) show a 
uniform
brightness distribution with no filamentary substructure typical of relic
sources.
At this high angular resolution the low brightness structure R1B to the 
west side of R1 is completely resolved; 
for this reason in Fig. \ref{fig:4} we show a smaller field of view.
 
The discrete sources imbedded in the extended radio source
are only a few and not relevant: a discrete source is present at  
the A3412 cluster
center (see Fig. \ref{fig:4}). 
This discrete radio source is located at the
X-ray peak emission and we identify it as the brightest cluster galaxy of 
A3412.
This source could be the origin of the compact X-ray emission, 
we note however that the 
radio power (see Sect. 4.3.3) is significantly lower than 3C radio sources
with a similar X-ray luminosity (see e.g. \cite{mas13}), and 
significantly lower than expected from 
the correlation between X-ray and radio luminosity discussed by 
\cite{bal06}.
Therefore we tentatively identify the X-ray emission present here, with the 
compact cluster A3412.
One more extended discrete radio source is at 
the periphery
of A3412 East-side, just outside the diffuse, non-thermal source 
(Fig. \ref{fig:4}). 
It could be a head-tail or double source according to the 
optical identification.

A few additional sub-mJy, slightly extended discrete sources are visible. 
All discrete sources have been subtracted before
measuring the radio parameters of the diffuse, non-thermal source.

\cite{vwe13} distinguish in the diffuse emission five elongated 
components (see their Fig. 3.Left), even if they classify this complex as
a radio relic. From the comparison of our Fig. \ref{fig:3} and Fig. \ref{fig:4}
we do not find a clear evidence of different sub-components, but
connected regions with a slightly higher and lower brightness, as expected
(and found) in many diffuse sources (halos and relics) correlated to the 
turbulence and/or shock waves in the hot ISM.
  
\subsection{Polarized emission}

\begin{figure}
   \centering
   \includegraphics[width=7.0cm]{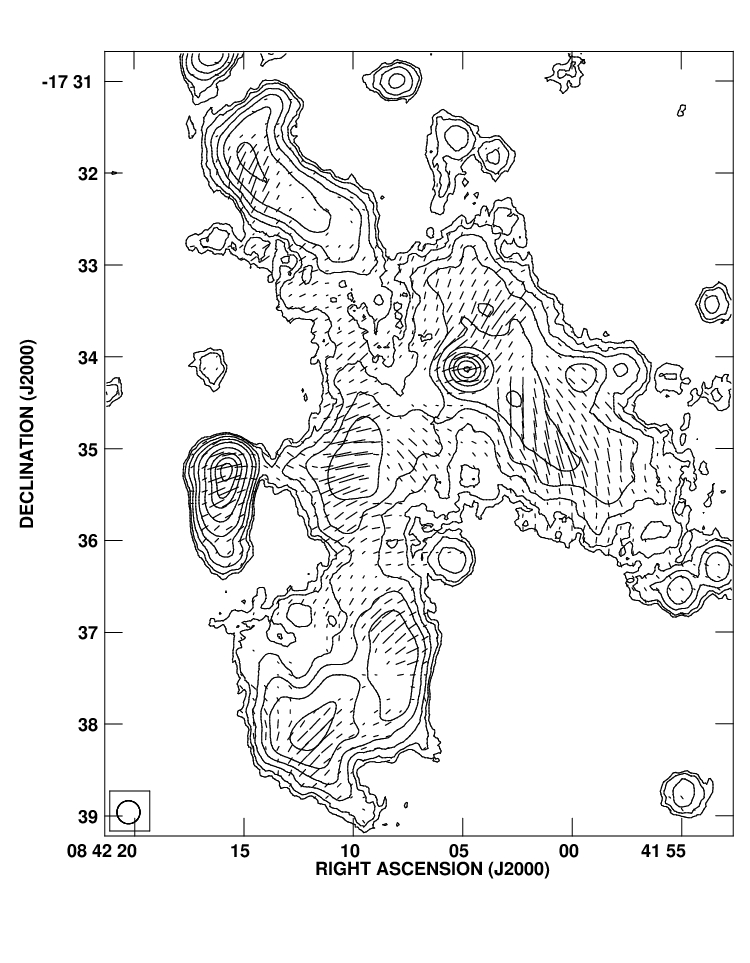}
      \caption{Contour image of the diffuse radio source (total intensity). 
The HPBW is 15''; levels are: 0.03 0.05 0.1 0.2 0.4 1 2 3 4 5
mJy/beam. Overimposed are vectors oriented as the polarization angle, with
a length proportional to the polarized intensity: 1'' = 1.25$\times$ 10$^{-5}$ 
Jy/beam.} 
         \label{fig:6}
\end{figure}

The extended diffuse, non thermal source is strongly polarized. 
In Fig. \ref{fig:6} we show
the total intensity image with superimposed polarization vectors
(no correction was applied to their orientation). In this 
figure the extreme region at SW (R1B) does not appear since no polarization is 
detected here. 

The average polarization percentage in R1 
is = 18\%; in the external R1B region we can give only an 
upper limit: $<$ 12\%. The R2 average polarization percentage is 17\%. 
The polarized vectors in R1 are oriented 
on the average at $\sim$ 20$^\circ$ in the SW region and at 
$\sim$ -25$^\circ$ in the NE region. In R2 vector orientation moves 
from $\sim$ 90$^\circ$ to -45$^\circ$. 

The polarization vectors show no preferential orientation. There are several
changes of direction on scales of 1-2 arcmin. This could suggest 
the presence of significant Rotation Measure (RM) effects,
due to a magnetized ICM. Being very distant from the A3411
center this ICM has to be related to the presence of A3412.

\subsection{Diffuse and discrete source parameters}
 
\subsubsection{The R1 \& R2 structures}

The total flux density of the R1+R2 structure is 67 $\pm$ 5 mJy (corresponding
to log P = 24.71 W/Hz), after the
subtraction of discrete sources. We note that: 1) the flux 
density uncertainty is not due to the noise or zero level in the image, but
to the problem to include or not some peripheral regions; 2) the only
relevant discrete sources are the one at the center of A3412, and 
the peripheral
extended source just outside the region R2. Other discrete sources are
sub-mJy sources. 
We have measured
the flux density in images with different angular resolutions and results
are consistent. Most of the flux density is in the R1 structure (considering 
also the R1B structure): 38 mJy (log P = 24.47 W/Hz), for R2 we measured 28 mJy
(log P = 24.34 W/Hz).
The total extension of R1 is $\sim$ 11' corresponding to $\sim$ 1.9 Mpc, and
$\sim$5' for R2 corresponding to $\sim$ 0.86 Mpc. As expected, these
values are in agreement within the uncertainties with \cite{vwe13}.

\subsubsection{The halo source in A3411}

In low resolution images a diffuse, non-thermal source is present 
coincident with the
central region of A3411. We identify this emission as
a diffuse small-size halo source. The total flux density from this region 
is $\sim$ 5 mJy
(in agreement with \cite{vwe13}), however after an accurate subtraction
of unrelated discrete sources selected in the high resolution image, we 
estimate a total flux density of the halo source = 
1.9 $\pm$ 0.3 mJy (corresponding to log P = 23.16 W/Hz). 
No polarized flux has been detected in this 
region.
We estimate a size of $\sim$ 2.5', corresponding to $\sim$ 430 kpc
(see Fig. \ref{fig:fighalo}).
This measure is very uncertain being the halo brightness at a few sigma level, 
and affected by confusion.
The estimated radio power, halo linear size and X-ray luminosity are in 
agreement with the X-ray luminosity-radio power and halo linear size-radio
power correlations (see \cite{fer12}).

\subsubsection{The two discrete sources in A3412}
 
The discrete source present at the center of A3412, shows a double 
morphology with a total flux density of 8.5 mJy (log P = 23.82 W/Hz),
and a size $\sim$
10''($\sim$ 30 kpc). It is not polarized, at a level of $<$ 5\%. 

The second discrete source at the periphery of R2, shows a double or head-tail 
morphology, with a total flux density = 12.1 mJy (log P = 23.97 W/Hz),
and a size $\sim$ 45'' ($\sim$ 120 kpc).
The average polarized flux percentage is $\sim$ 7\%. 

\section{Discussion}

\subsection{The giant diffuse non-thermal source}

The non-thermal properties of this region are very complex.
\cite{vwe13} identify the complex diffuse radio emission as a radio
relic broken into five fragments suggesting that the shock responsible
for the radio emission has been broken up due to interaction with a 
large-scale galaxy filament connected to A3411 or other substructures.

Observational data presented here, suggest that the 
elongated structure R1 is a diffuse non-thermal source associated to A3411, 
with many properties similar to relic sources: elongated 
shape and polarization
properties. This interpretation is supported by the presence of a secondary
peak in the X-Ray emission of A3411 confirming that R1 is located outward
of a merging structure.
However, the origin of the extended feature R2 
is puzzling. Although the radio properties 
(elongated shape and polarization) suggest its 
identification as a relic source too, its extension aligned with the 
A3411 -- A3412 structure as visible in the X-ray image, requires a more 
complex dynamical scenario. We remember that most (all) known 
elongated relics show the major axis roughly perpendicular to the direction
of the cluster centers (see e.g. \cite{bru11}).

Some radio properties of these sources  
are consistent with properties of relics (see e.g. \cite{fer12})
and in agreement with known correlations 
(e.g., radio power versus linear size, and radio 
power versus X-ray Luminosity).
R1 would be one of the largest relic sources.
However we note that 
R1 and R2 are diffuse complex radio sources which are clearly connected and 
thus of similar origin and nature.
We note also that, 
when observed at high resolution,
R1 and R2 are diffuse,
homogeneous regions with no evidence of substructures. There is no sharp edge
in the brightness distribution, expected if the radio emission is the tracer
of shocks in merger events.
Shock waves are expected to accelerate radio emitting electrons, and to 
amplify and order local magnetic fields. 
In R1 and R2 we have a highly polarized emission with vectors locally 
ordered but changing orientation on scales of 1-2 arcmin. 

The morphology and properties of the 
diffuse non-thermal source can be explained by the presence of 
a complex condition with accretion shocks able to accelerate
electrons, combined with 
turbulence in the thermal gas connected to the presence of A3412. 
The multiple mergers of the A3411 -- A3412 structure could be 
the origin of the R1+R2 diffuse source, 
with R2 oriented along the giant filament merging into A3411.
The R2 structure could be powered by accretion shocks as material 
falls onto the filament, as e.g., suggested by \cite{bro11} for the relic 
present in the Coma cluster. We remember that in the Coma cluster hosting
a radio halo and a relic source, also a Mpc scale radio bridge 
connecting 
the halo to the relic and elongated in the same direction of the giant 
Coma-A1367 supercluster (\cite{kim89}) is present.

Alternatively \cite{vwe13} suggest that the complex morphology reflects the 
presence of electrons in fossil radio bubbles that are re-accelerated by a 
shock. In their Fig. 3 five regions are individuated. 
We think unlikely this hipothesis because the presence of so many 
bubbles implies a strong past activity of some radio galaxies that at present
should be completely radio quiet. No connection with the present 2 radio 
galaxies in A3412 is visible. Moreover as already discussed, at high resolution
the morphology of the giant non-thermal source is very homogeneous with
no evidence of shocks and at high and low resolution the whole structure
is connected and likey to have a common origin.

New multifrequency observations to study the
spectral index distribution and polarization properties are necessary  to 
properly discuss these structures. The presence of one more optical 
condensation to the North of A3411 (see Table 1), suggests a possible 
more complex
structure even if present data do not show radio or X-ray emission from this
region.

\subsection{The central Halo source in A3411}

A diffuse, non-thermal source at the center of A3411 is
present in our images. Despite of the large uncertainties previously
discussed, we note that 
in comparison with other radio halos, the halo in A3411 is one with 
the lowest radio power.
Only the A3562 halo (\cite{ven03}) is known so far to show a lower 
radio power
(\cite{fer12} and references therein). 

These faint radio halos both follow the general
correlations between X-ray luminosity, radio power and linear size 
for radio halos
(see e.g \cite{fer12}). It is important to increase the statistical 
information on these low power radio halos to discuss if all merging 
clusters host a central diffuse source (in most cases not visible with
present radio telescopes because of sensitivity, see e.g. \cite{xu12}) 
or if an unknown critical value of the halo radio power is present.

\section{Conclusions}

We have presented here the detection of a complex non-thermal source
in the A3411-A3412 structure.
This source has been found at the periphery
of A3411, where the cluster A3412 is located, aligned with the 
A3411 merging structure. 
We suggest that this giant, diffuse, and peculiar
non-thermal source is related  to a filamentary  
structure and to the presence of multiple mergers.
Better optical, radio and X-ray data are necessary to investigate this
complex region and to derive its physical properties. 
A few cases are present in literature where intergalactic
filaments have been firmly detected and only in the Coma cluster 
(\cite{kim89}), and in ZW2341.1+0000 (\cite{bag02} and \cite{gio10})
a non-thermal emission has been found. 
To increase the number of these
sources is important to understand physical conditions in forming cosmological
structures. We expect that the new generation of radio telescopes (SKA and
SKA-precursors) will increase the detection of these diffuse sources allowing
to extend our knowledge of non-thermal emission from clusters to filaments.

Moreover, we have found that the radio halo at the center of A3411 has a 
low power,
despite of the large uncertainties related to its low brightness.
It is important to
increase the number of known low power radio halos to confirm or not that they 
follow the general correlation between X-ray luminosity and radio power 
of Mpc--scale radio halos.

The presence of this source is related to an active merger well evident in the 
X-ray image. Data from the new generation of radio telescopes are needed to 
better study low power radio halos.

\section*{Acknowledgements}
The National Radio Astronomy
Observatory is operated by Ass. Univ., Inc., under cooperative
agreement with the National Science Foundation.
This research has made use of the NASA/IPAC Extragalactic Database (NED) 
which is operated by the Jet Propulsion Laboratory, California Institute of 
Technology, under contract with the National Aeronautics and Space 
Administration. We thank the Referee and the Scientific Editor for useful
suggestions and comments.

\label{lastpage}

\end{document}